\documentstyle[12pt, psfig]{article}
\newcommand{\mysection}{\setcounter{equation}{0}\section}

\def\beq{\begin{equation}}
\def\eeq{\end{equation}}
\def\beqa{\begin{eqnarray}}
\def\eeqa{\end{eqnarray}}

\begin{document}

\begin {flushright}
EDINBURGH 97/14
\end {flushright} 
\vspace{3mm}
\begin{center}
{\Large \bf Resummation of next-to-leading logarithms in top quark production
\footnote{Presented at the QCD 97 Euroconference, 
Montpellier, July 3-9, 1997}} 
\end{center}
\vspace{2mm}
\begin{center}
Nikolaos Kidonakis\footnote{This work was supported by the PPARC 
under grant GR/K54601}\\
\vspace{2mm}
{\it Department of Physics and Astronomy\\
University of Edinburgh\\
Edinburgh EH9 3JZ, Scotland} \\
\vspace{2mm}
August 1997
\end{center}

\begin{abstract}
We discuss the resummation of next-to-leading logarithms (NLL)
in heavy quark production near threshold. 
Results are presented for top quark production at the Fermilab Tevatron.
\end{abstract}

\mysection{Introduction}
There has been a lot of interest recently in the resummation
of soft gluon radiation near threshold for heavy quark production 
in hadronic collisions [1-4]. This resummation is a
direct consequence of the factorization of short- from long-distance
effects in QCD scattering [5]. In the perturbative expansion for the heavy 
quark cross section one finds
logarithmic terms that can be 
resummed to all orders of perturbation theory.
These logarithms come from distributions that are singular for $z=1$, 
where $z=Q^2/s$, with $Q^2$ and $s$ the invariant
mass squared of the produced heavy quark pair and the partons
in the incoming hadrons, respectively.
One of the interesting features of singular distributions
in QCD-induced cross sections is their sensitive dependence
on the color exchange in the hard scattering.
This feature as well as the presence of final state interactions
are the main complications relative to the Drell-Yan process [6].
We will show that these effects contribute only at NLL.
As we discuss below, resummation is conveniently carried out in terms
of moments, and logarithmic dependence on the moment variable exponentiates.
The moments have to be inverted to derive the physical cross section.
Previous formalisms [1-4] resum only leading logs 
(and some NLL) and there 
has been a debate as to which subleading logs to keep. In our analysis
(in moment space) we have for the first time resummed all the NLL so we have 
effectively pushed the debate in [1-3] to the next-to-next-to-leading 
logarithmic level. In Section 4 we invert moments using the method 
in Ref. 1. 

\mysection{Resummation for QCD hard scattering}
For the hadronic process
$h_1(p_1)+h_2(p_2) \rightarrow Q \bar{Q}+X$ the cross section $\sigma$, 
can be written as a convolution of parton distribution functions $\phi$, 
with the partonic hard scattering $\hat \sigma$,
\beq
\sigma_{h_1h_2\rightarrow Q{\bar Q}}=\sum_{f=q, {\bar q}, g}
\phi_{f/h_1}\otimes\hat{\sigma}_{f{\bar f}\rightarrow Q{\bar Q}} 
\otimes \phi_{{\bar f}/h_2}\, .
\eeq
This convolution becomes a product in terms of moments,
\begin{equation}
\tilde \sigma_{h_1 h_2 \rightarrow Q \bar Q}(N)
=\sum_{f=q, {\bar q}, g} \tilde \phi_{f/h_1}(N) \: 
\hat \sigma_{f \bar f \rightarrow Q \bar Q}(N) \:
\tilde \phi_{\bar f/h_2}(N) \, ,
\end{equation}
where $\tilde \sigma(N)=\int_0^1 d\tau \; \tau^{N-1} \; \sigma(\tau),
\quad \quad \hat \sigma(N)=\int_0^1 dz \; z^{N-1} \; \hat \sigma(z)$, and
$\tilde \phi(N)=\int_0^1 dx \; x^{N-1} \; \phi(x)$.
Here $\tau=Q^2/S$ where $S$ is the invariant mass squared 
of the incoming hadrons.
Singular distributions in $z$ give rise under moments to 
logarithmic $N$-dependence, as in
\beq
\int_0^1 dz\; z^{N-1}\; 
\left [ {\ln^m(1-z)\over 1-z} \right ]_+
\propto \ln^{m+1}N . 
\eeq
Thus, the large-$N$ behavior is a 
diagnostic for singular distributions in $1-z$.
We can derive the resummation of soft gluon contributions 
by reexpressing  moments of the cross section (with $h_1=f$, $h_2=\bar f$)
in a form in which collinear gluons are factorized
into alternate parton distributions $\psi$, and soft gluons into a
function $S_{JI}$,
\beq
\sigma_{f{\bar f}\rightarrow Q{\bar Q}}
=
\psi_{f/f}\; \otimes\; h_J^*h_I\; \otimes\; S_{JI}\; \otimes\; 
\psi_{{\bar f}/{\bar f}}\, .
\eeq
This factorization takes into account the color exchange
at the hard scatterings, $h_I$ and $h^*_J$ in the amplitude and its complex
conjugate, respectively, where $I$ and $J$ are color tensor indices.  

From Eqs. (2.1) and (2.4) we find
$\hat \sigma(N)=[\psi(N)/\phi(N)]^2 \, h_J^* \, S_{JI}(N) \, h_I$.
The ratio of the universal $\psi$ and $\phi$ has been analyzed in
Drell-Yan production \cite{DY}.

The soft matrix $S_{JI}(Q/(N\mu))$ requires renormalization:
$S_{JI}^{(0)}=(Z_S^\dagger)_{JB}  S_{BA}$ 

\noindent $ \times Z_{S,AI}$,
where $Z_{S,JI}$ is a matrix of renormalization constants \cite{BottsSt}.
Hence $S_{JI}$ satisfies the
renormalization group equation
\begin{equation}
\mu \frac{dS_{JI}}{d\mu}
=-(\Gamma^\dagger_S)_{JB}S_{BI}-S_{JA}(\Gamma_S)_{AI}\, .
\end{equation}
In a minimal subtraction scheme, 
and with $\epsilon=4-n$, the anomalous dimension matrix $\Gamma_S$ is 
\begin{equation}
\Gamma_S (g)=-\frac{g}{2} \frac {\partial}{\partial g}{\rm Res}_{\epsilon 
\rightarrow 0} Z_S (g, \epsilon).
\end{equation}
At the level of leading logarithms of $N$
in $S_{JI}$, and therefore at NLL of $N$ in
the cross section as a whole, we 
choose a color basis in which the anomalous dimension matrix is
diagonal, with eigenvalues $\lambda_I$ for each basis color
tensor labelled by $I$. Then, the solution to Eq. (2.5) is 
\begin{eqnarray}
\tilde S_{JI}\left(\frac{1}{N}, \, \alpha_s(Q^2)\right)&=&\tilde
S_{JI}\left(1, \, \alpha_s\left(\left[\frac{Q}{N}\right]^2\right)\right)  
\nonumber \\ && \! \! \! \!
\times \exp\left[-\int^{Q}_{Q/N}\frac{d \bar{\mu}}{\bar{\mu}}
[\lambda_I(\alpha_s(\bar{\mu^2}))
+\lambda^*_J(\alpha_s(\bar{\mu^2}))]\right]. 
\end{eqnarray}
Thus, we find for the partonic cross section [8,9]
\beq
{\hat \sigma}_{f{\bar f}\rightarrow Q{\bar Q}}(N)
= A'  e^{E_{JI}(N,\theta,Q^2)}  h^*_J \! \left(1,\alpha_s(Q^2)\right)  
{\tilde S}_{JI} \! \left(1,\alpha_s
\left(\left[\frac{Q}{N}\right]^2\right)\right)
h_I \!\left(1, \alpha_s(Q^2)\right) 
\eeq
where $A'$ is an overall constant and $\theta$ is the
center-of-mass scattering angle. The exponent is
\beqa
E_{JI}(N,\theta,Q^2)&=&E_{\rm DY}(N, Q^2) 
- \int_0^1 dz \frac{z^{N-1}-1}{1-z}
\left[ g_3^{(I)}[\alpha_s((1-z)^2 Q^2),\theta] \right.
\nonumber \\ && \quad \quad \quad \quad \quad \quad \quad \quad \quad 
\left. \mbox{}+g_3^{(J)*}[\alpha_s((1-z)^2 Q^2),\theta]\,\right]\, ,
\eeqa
where $E_{\rm DY}$ is the Drell-Yan exponent, and
\begin{equation} 
g_3^{(I)}[\alpha_s,\theta]=-\lambda_I[\alpha_s,\theta]
+\frac{\alpha_s}{\pi} C_{F, A} \, .
\end{equation}

\mysection{The $\Gamma_S$ matrices for $q \bar{q} \rightarrow Q \bar{Q}$
and $gg \rightarrow Q \bar{Q}$}

We now give explicit results for the anomalous dimension matrices $\Gamma_S$.
We begin with quark-antiquark annihilation,
$q(p_a)+{\bar q}(p_b) \rightarrow {\bar Q}(p_1) + Q(p_2)$.
The explicit calculation is given in \cite{Thesis}.
Here we give the results for the anomalous dimension matrix in a color
tensor basis consisting of singlet and octet exchange in the $s$ channel,
\beqa
c_1=\delta_{ab} \, \delta_{12}, \quad \quad 
c_2=-\frac{1}{2N} c_1 +\frac{1}{2} \delta_{a2} \, \delta_{b1} \, .
\eeqa
The anomalous dimension matrix (shifted by $C_F \alpha_s/\pi$ as in 
Eq. (2.10)) in this color basis and in an axial gauge $A^0=0$ 
is \cite{Thesis}
\begin{eqnarray}
\Gamma_{S, 11}&=&-\frac{\alpha_s}{\pi}C_F(L_{\beta}+1+\pi i),
\nonumber \\
\Gamma_{S, 21}&=&\frac{2\alpha_s}{\pi}
\ln\left(\frac{u_1}{t_1}\right), \quad \quad
\Gamma_{S, 12}=\frac{C_F}{2C_A}\Gamma_{S, 21},
\nonumber \\
\Gamma_{S, 22}&=&\frac{\alpha_s}{\pi}\left\{C_F
\left[4\ln\left(\frac{u_1}{t_1}\right)-L_{\beta}-1-\pi i\right]\right.
\nonumber \\ && \! \! \! \! \! \! \! \! \! \! \! \! \! \!
\left.\mbox{}+\frac{C_A}{2}\left[-\ln\left(\frac{u_1^2 m^2 s}{t_1^4}\right)
+L_{\beta}+\pi i \right]\right\}\, ,
\end{eqnarray}
where $s=(p_a+p_b)^2$, $t_1=(p_a-p_1)^2-m^2$, $u_1=(p_b-p_1)^2-m^2$,
and $m$ is the heavy quark mass.
Also
\begin{equation}
L_{\beta}=\frac{1-2m^2/s}{\beta}\left(\ln\frac{1-\beta}{1+\beta}
+\pi i \right)\, ,
\end{equation}
where $\beta=\sqrt{1-4m^2/s}$.
$\Gamma_S$ is diagonalized in this singlet-octet basis
for arbitrary $\beta$ when the scattering angle is
$\theta=90^\circ$ (where $u_1=t_1$).
It is also diagonalized at $\beta=0$.

Next we discuss gluon fusion,
$g(p_a)+g(p_b) \rightarrow {\bar Q}(p_1) + Q(p_2)$.
We choose the following color basis:
\begin{equation}
c_1=\delta^{ab}\,\delta_{21}, \; c_2=d^{abc}\,T^c_{21},
\; c_3=i f^{abc}\,T^c_{21}.
\end{equation}
Then in the $A^0=0$ gauge the anomalous dimension matrix 
(shifted by $C_A \alpha_s/\pi$ as in Eq. (2.10)) is \cite{Thesis}
\begin{eqnarray}
\Gamma_{S,11}&=&\frac{\alpha_s}{\pi}[-C_F(L_{\beta}+1)-C_A\pi i],
\nonumber \\ 
\Gamma_{S,21}&=&0, \quad \quad \Gamma_{S,12}=0,
\nonumber \\ 
\Gamma_{S,31}&=&2\frac{\alpha_s}{\pi}\ln\left(\frac{u_1}{t_1}\right),
\nonumber \\
\Gamma_{S,22}&=&\frac{\alpha_s}{\pi}\left\{-C_F(L_{\beta}+1) \right.
\nonumber \\ &&
\left.\mbox{}+\frac{C_A}{2}\left[\ln\left(\frac{t_1 u_1}{m^2 s}\right)
+L_{\beta}-\pi i \right]\right\},
\nonumber \\
\Gamma_{S,32}&=&\frac{N^2-4}{4N}\Gamma_{S,31}, \quad \quad
\Gamma_{S,13}=\frac{1}{2}\Gamma_{S,31},
\nonumber \\
\Gamma_{S,23}&=&\frac{C_A}{4}\Gamma_{S,31}, \quad \quad
\Gamma_{S,33}=\Gamma_{S,22}.
\end{eqnarray}
We note that the matrix is diagonalized at $\theta=90^\circ$
and also at $\beta=0$.

We have checked that the one-loop expansion of our results for both channels 
are consistent with [10].

\mysection{Results for top quark production at the Fermilab Tevatron}
We are now interested in the magnitude
of the NLL terms, particularly the $g_3$ contribution,
relative to the LL results of \cite{LSN}.
Therefore we use a similar cutoff scheme, modifying the definitions of 
the exponent in Eq. (2.9) to work directly in momentum space.
This method uses the correspondence between logarithms of the
moment variable $N$ and logarithmic terms in the momentum space variable 
$1-z$. We work at $\theta=90^\circ$ to avoid the diagonalization of 
the anomalous dimension matrices.

\begin{figure}
\centerline{\psfig{file=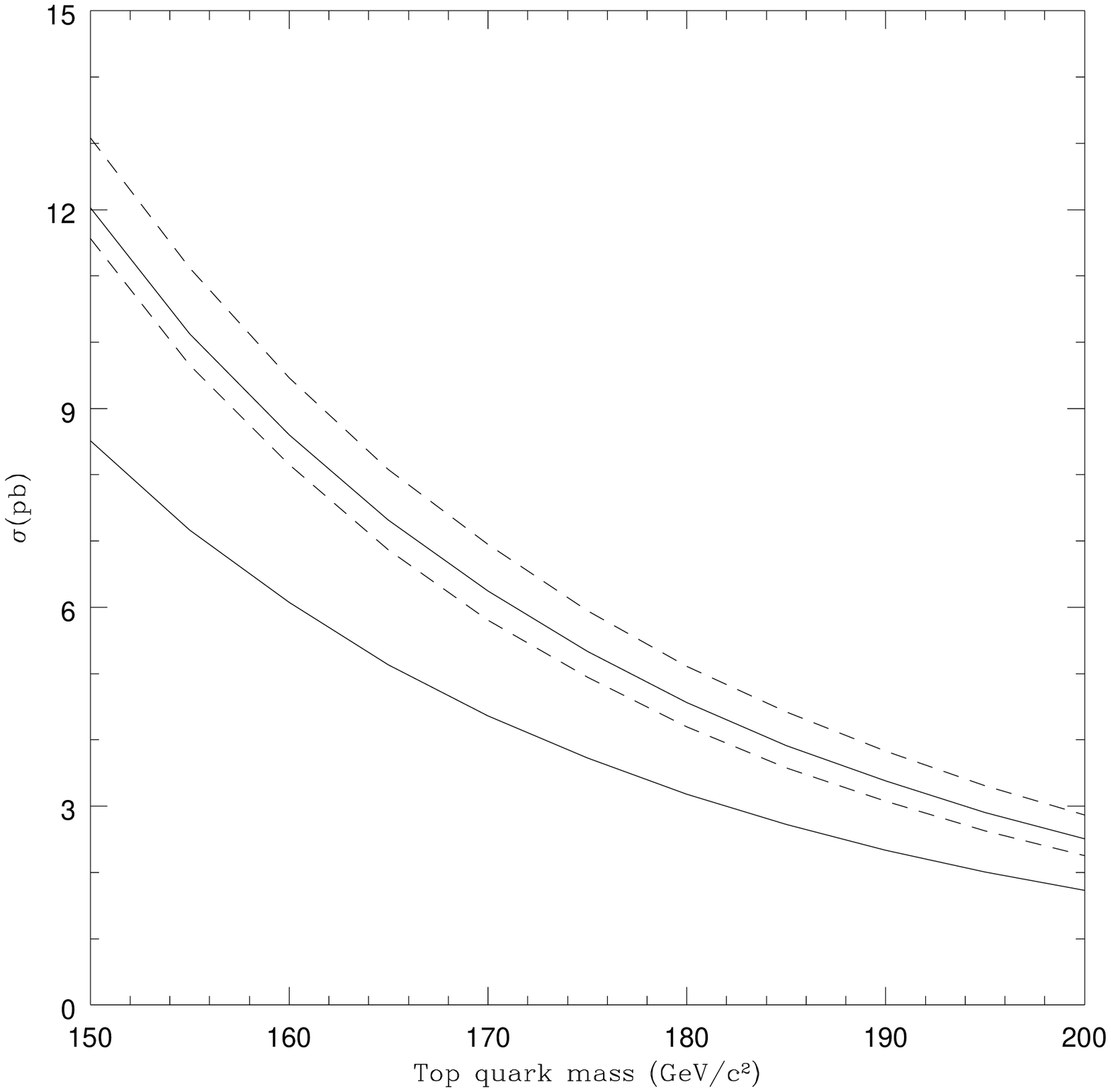,height=12cm,width=12cm,clip=}}
{Fig. 1. The resummed top quark production cross section
at $\theta=90^\circ$ in the $\overline {\rm MS}$ scheme. 
We show the resummed results without the $g_3$ terms (lower solid line)
and the resummed results with all the NLL terms ($Q=m$, upper solid line;
$Q=2m$, lower dashed line; $Q=m/2$, upper dashed line).}
\label{fig. 1}
\end{figure}
The resummed partonic cross section is \cite{nkjsrv}
\beq
\sigma_{ab}(s, m^2)=-\int_{s_{\rm cut}}^{s-2ms^{1/2}} ds_4 \;
f_{ab}\left(\frac{s_4}{2m^2}\right) \;
\frac{d \overline{\sigma}_{ab}^{(0)}(s, s_4, m^2)}{ds_4}\, ,
\eeq
where $ab = q \overline q$ or $gg$ and $s_4=2m^2 \; (1-z)$ at threshold.    
$d \overline{\sigma}_{ab}^{(0)}/ds_4$ is the differential of the
Born cross section \cite{LSN}. 
After replacing $z^{N-1}-1$ in Eq. (2.9) by $-1$ and
introducing $\omega=1-z$, the exponential function for the $q \bar q$
channel is \cite{nkjsrv} 
\begin{equation}
f_{q \bar q}\left( \frac{s_4}{2m^2} \right) =
\exp [E_{q \bar q}^{\rm DY}
+ E_{q \bar q}(\lambda_{\rm octet}) ]\,,
\end{equation}
where $E_{q \bar q}^{\rm DY}$ is the Drell-Yan part.
The color-dependent $g_3$ contribution in Eq.\ (2.9) leads to
\beq 
E_{q \bar q}(\lambda_i)=
-\int_{\omega_0}^1\frac{d\omega}{\omega}
\left\{\lambda_i\left[\alpha_s \left(\frac{\omega^2 Q^2}{\Lambda^2}\right),
\theta=90^\circ\right]+\lambda_i^*
\left[\alpha_s \left(\frac{\omega^2 Q^2}{\Lambda^2}\right),\theta=90^\circ
\right]\, \right\} 
\eeq
where $i$ denotes singlet or octet.
Since our calculation is not done in moment space, the $\omega$ integral
is cut off at $\omega_0= s_4/2m^2$.  Because the running coupling
constant diverges when $\omega^2 Q^2/\Lambda^2 \sim 1$,
the minimum cutoff in eq.\ (4.3)
is $s_{\rm cut} = s_{4, {\rm min}} \sim 2 m^2 \Lambda/Q$,
where $\Lambda$ is the QCD scale parameter. 
In general we choose a larger value for the cutoff consistent with the sum 
of the first few terms in the perturbative expansion; here 
we cut off the $z$-integration at $z=1-10\Lambda/Q$.

The treatment of the gluon-gluon channel
in the $\overline {\rm MS}$ scheme is very similar
but now we have three distinct color structures.  However,
only two of them are independent.  We define
$f_{gg}$ for each eigenvalue so that
$f_{gg,i} = \exp[E_{gg}^{\rm DY}
+ E_{gg}(\lambda_i)]$, in complete analogy with the relations for the
$q \bar q$ channel.
For details see Ref. 11.

In Fig. 1 we show results 
for the resummed top quark cross section at the Fermilab Tevatron with
$\sqrt{S}=1.8$ TeV (with and without the NLL $g_3$
terms) versus the top quark mass. We used the MRSD$-'$ parton densities
[12]. We also show the variation of the cross section with 
the factorization scale.
Then for $m=175$ GeV/$c^2$ and $Q=m$ the total $t {\bar t}$ resummed NLL
$\overline{\rm MS}$ cross section at $\theta=90^\circ$ is 5.3 pb.
Without the NLL $g_3$ contribution the cross section is 3.7 pb.  
Thus the NLL $g_3$ terms enhance the total $\overline{\rm MS}$
cross section considerably.

\mysection{Conclusions}
We have given explicit results for the resummation of next-to-leading 
logarithms in top quark production. 
We have shown that the NLL terms are numerically significant.
Our methods have been applied to $b$-quark production at HERA-B,
with similar conclusions \cite{nkjsrv}, 
and they can also be applied to jet production \cite{kos}.
We have recently completed the evaluation of the
anomalous dimension matrices for $q \bar q \rightarrow gg$,
$qg \rightarrow qg$, and the more complicated case of $gg \rightarrow gg$,
all relevant in jet production \cite{kos}.

\mysection*{Acknowledgements}
The work in Sections 2 and 3 was done in collaboration with George Sterman.
The work in Section 4 was done in collaboration with Jack Smith and Ramona 
Vogt. I would like to thank them for their help and useful conversations.

\mysection*{Questions}

\noindent {\it L. Trentadue, Parma:}

You have shown some numerical outputs of your calculation. 
You have used the method of Laenen, Smith, and van Neerven to cut-off
the $z \rightarrow 1$ limit. Could not the results be affected by 
this by hand limiting of the final phase space?

\vspace{1mm}

\noindent {\it N. Kidonakis:}

We were interested in the relative size of the NLL. Of course the result 
depends somewhat on the procedure used and the value of the cutoff. 
However, we calculated the cross section with different
cutoffs and found that the NLL contribution is always significant, 
and that is our main conclusion.

\end{document}